# 後量子密碼學憑證混合方案比較研究

Abel C. H. Chen
Information & Communications Security Laboratory,
Chunghwa Telecom Laboratories
Email: chchen.scholar@gmail.com
ORCID: 0000-0003-3628-3033

## 摘要

隨著量子計算硬體技術的成熟,搭配量子演算法將有可能在多項式時間內破解 RSA 密碼學和橢圓曲線密碼學。有鑑於此,美國國家標準暨技術研究院(National Institute of Standards and Technology, NIST)於2024年8月陸續制定後量子密碼學標準,並且已經規劃後量子密碼學遷移時程。因此,在憑證管理系統應用規劃符合後量子密碼學標準的 X.509憑證是現階段重要的議題之一。除此之外,為了提升安全性和考慮遷移議題,目前國際上已經在 X.509憑證基礎上提出複合(Composite)混合憑證方案、催化劑(Catalyst)混合憑證方案、變色龍(Chameleon)混合憑證方案等。本研究深入討論與比較複合混合憑證方案、催化劑混合憑證方案、變色龍混合憑證方案,並且從憑證長度、計算時間長度、以及遷移議題等面向來比較各種混合憑證方案,以評估各種混合憑證方案適用的應用與服務。

**關鍵詞**:後量子密碼學、X.509憑證、複合(Composite)混合憑證方案、催化劑(Catalyst)混合憑證方案、變色龍(Chameleon)混合憑證方案。

# 後量子密碼學憑證混合方案比較研究

## 摘要

隨著量子計算硬體技術的成熟，搭配量子演算法將有可能在多項式時間內破解 RSA 密碼學和橢圓曲線密碼學。有鑑於此，美國國家標準暨技術研究院(National Institute of Standards and Technology, NIST)於2024年8月陸續制定後量子密碼學標準，並且已經規劃後量子密碼學遷移時程。因此，在憑證管理系統應用規劃符合後量子密碼學標準的 X.509憑證是現階段重要的議題之一。除此之外，為了提升安全性和考慮遷移議題，目前國際上已經在 X.509憑證基礎上提出複合(Composite)混合憑證方案、催化劑(Catalyst)混合憑證方案、變色龍(Chameleon)混合憑證方案等。本研究深入討論與比較複合混合憑證方案、催化劑混合憑證方案、變色龍混合憑證方案，並且從憑證長度、計算時間長度、以及遷移議題等面向來比較各種混合憑證方案，以評估各種混合憑證方案適用的應用與服務。

**關鍵詞**：後量子密碼學、X.509憑證、複合(Composite)混合憑證方案、催化劑(Catalyst)混合憑證方案、變色龍(Chameleon)混合憑證方案。

## 1. 前言

近幾年量子計算軟硬體技術快速成長，已經逐漸成為可以用來解決許多困難問題的重要工具。其中，硬體技術方面，Google 在 2024 年於《Nature》期刊上發表論文 "Quantum Error Correction Below the Surface Code Threshold" 可以做到大幅降低量子計算錯誤率[1]；軟體技術方面，Shor 在 1997 年於《SIAM Journal on Computing》期刊上發表論文 "Polynomial-time Algorithms for Prime Factorization and Discrete Logarithms on a Quantum Computer" 可以做到在多項式時間內分解質因數和解決離散對數問題[2]。在量子計算軟硬體技術的搭配下，將有可能破解建構在質因數問題的 RSA 密碼學和建構在離散對數問題的橢圓曲線密碼學[3]。因此，2024 年 8 月美國國家標準暨技術研究院(National Institute of Standards and Technology, NIST)陸續出版多項後量子密碼學(Post-Quantum Cryptography, PQC)相關的聯邦資訊處理標準(Federal Information Processing Standards, FIPS)，制定了一項金鑰封裝機制標準(基於模格金鑰封裝機制(Module-Lattice-Based Key-Encapsulation Mechanism, ML-KEM)[4])和兩項數位簽章標準(基於模格數位簽章演算法(Module-Lattice-Based Digital Signature Algorithm, ML-DSA)[5]、基於無狀態雜湊數位簽章演算法(Stateless Hash-Based Digital Signature Algorithm, SLH-DSA)[6])，以抵抗來自量子計算攻擊的安全威脅。並且，目前還有兩個待標準化的演算法，包含有目前數位簽章演算法 Falcon [7]和金鑰封裝機制 HQC (Hamming Quasi-Cyclic)[8]。除此之外，目前仍持續進行第二輪遴選額外數位簽章演算法，預期將會再評選出短簽章的數位簽章演算法[9]。

不同的標準演算法，在金鑰長度、密文長度、簽章長度、計算時間長度、以及安全的困難數學問題等不同的面向都有不同表現，未來將可以根據應用和服務需要選擇合適的標準演算法。

美國國家標準暨技術研究院有鑑於未來十年量子計算技術將可能成熟，所以在 2024 年已經有擬定後量子密碼學遷移草案，並且草案文件中指出將於 2035 年全面禁用 RSA 密碼學和橢圓曲線密碼學[10]。另外，由於目前已經制定了數位簽章演算法標準，包含基於模格數位簽章演算法、基於無狀態雜湊數位簽章演算法，建議根據表 1 描述之演算法和參數來部署；並且目前已經制定了數位封裝演算法標準，包含基於模格金鑰封裝機制，建議根據表 2 描述之演算法和參數來部署。

**表 1 後量子密碼學數位簽章演算法標準[5], [6]**

| 密碼學類別 | 數位簽章演算法及其參數 | 安全等級 |
|---|---|---|
| 基於格密碼學 | ML-DSA-44 | 2 |
| | ML-DSA-65 | 3 |
| | ML-DSA-87 | 5 |
| 基於雜湊密碼學 | SLH-DSA-SHA2-128s<br>SLH-DSA-SHAKE-128s<br>SLH-DSA-SHA2-128f<br>SLH-DSA-SHAKE-128f | 1 |
| | SLH-DSA-SHA2-192s<br>SLH-DSA-SHAKE-192s<br>SLH-DSA-SHA2-192f<br>SLH-DSA-SHAKE-192f | 3 |
| | SLH-DSA-SHA2-256s<br>SLH-DSA-SHAKE-256s<br>SLH-DSA-SHA2-256f<br>SLH-DSA-SHAKE-256f | 5 |

**表 2 後量子密碼學金鑰封裝機制標準[4]**

| 密碼學類別 | 數位簽章演算法及其參數 | 安全等級 |
|---|---|---|
| 基於格密碼學 | ML-KEM-512 | 1 |
| | ML-KEM-768 | 3 |
| | ML-KEM-1024 | 5 |

有鑑於後量子密碼學標準化的發展，在網際網路工程任務組(Internet Engineering Task Force, IETF)也開始著手制定相關請求意見稿(Request for Comments, RFC)草案，包含後量子密碼學 X.509 憑證標準[11], [12]、後量子密碼學 X.509 混合憑證標準[13]-[16]、以及傳輸層安全性協定(Transport Layer Security, TLS)標準[17]，作為公開金鑰基礎建設(Public Key Infrastructure, PKI)遷移到後量子密碼學的重要參考之一。因此，本研究主要將聚焦在後量子密碼學 X.509 憑證標準和後量子密碼學 X.509 混合憑證標準，並且討論複合(Composite)混合憑證方案、催化劑(Catalyst)混合憑證方案、變色龍(Chameleon)混合憑證方案的設計原理，以及從憑證長度、計算時間長度、以及遷移議題等面向來比較各種混合憑證方案。

本文主要分為七個章節。第 2 節介紹後量子密碼學 X.509 憑證標準。第 3 節~第 5 節分別說明論複合混合憑證方案、催化劑混合憑證方案、以及變色龍混合憑證方案。第 6 節對各種混合憑證方案

進入深入的討論與比較。最後，第 7 節總節本研究的主要貢獻和未來發展方向。

## 2. X.509憑證標準及其格式

本節介紹後量子密碼學 X.509 憑證標準，並且分別表述支援數位簽章用的後量子密碼學 X.509 憑證[11]和支援金鑰封裝用的後量子密碼學 X.509 憑證[12]。值得注意的是，雖然這兩個技術目前是草案階段，但已經進入第 10 版以上，將有可能成為標準。本節案例，假設憑證擁有者是 Alice，並且 Alice 的數位簽章演算法採用 ML-DSA-44、金鑰封裝機制採用 ML-KEM-512，以及憑證中心(Certificate Authority, CA)的數位簽章演算法採用 ML-DSA-44。另外，本節僅是一個簡單示意，在實際部署上可以考慮採用更安全等級的標準演算法及其參數。

以支援數位簽章用的後量子密碼學 X.509 憑證(如圖 1 所示)為例，當憑證中心從註冊中心(Registration Authority, RA)收到 Alice 支援數位簽章用的憑證簽章請求(Certificate Signing Request, CSR)，並且確認內容正確後可以簽發支援數位簽章用的後量子密碼學 X.509 憑證給 Alice。在主題(Subject)的常用名稱(Common Name, CN)欄位中描述 Alice，並且設定主題公鑰資訊(Subject Public Key Information, SPKI)的 key 欄位設定為 Alice 的 ML-DSA-44 公鑰和 alg 欄位設定為 ML-DSA-44 的物件識別碼(Object Identifier, OID)。並且憑證將還包含序號、效期等資訊，也應一併包含，但本研究主要聚焦在後量子密碼學描述，所以該些欄位未進行深入描述。另外，由於此憑證主要作為數位簽章用，所以在擴展(Extensions)的金鑰用途(Key Usage)欄位中加入 digitalSignature。上述內容作為待簽章憑證(To-Be-Signed-Certificate, TBS-Certificate)，再由憑證中心運用憑證中心 ML-DSA-44 私鑰對為待簽章憑證產製 ML-DSA-44 簽章放到 Sig 欄位。

| Pure Certificate (X.509) |
|---|
| Subject: cn=Alice |
| SPKI: {<br>  alg: **ML-DSA-44 OID**<br>  key: **ML-DSA-44 key**<br>} |
| Extensions: |
| Key Usage: {digitalSignature} |
| Sig: {**ML-DSA-44 signature**} |

**圖 1 支援數位簽章用的後量子密碼學 X.509憑證**

以支援金鑰封裝用的後量子密碼學 X.509 憑證(如圖 2 所示)為例，當憑證中心從註冊中心收到 Alice 支援金鑰封裝用的憑證簽章請求，並且確認內容正確後可以簽發支援金鑰封裝用的後量子密碼學 X.509 憑證給 Alice。在主題的常用名稱欄位中描述 Alice，並且設定主題公鑰資訊的 key 欄位設定為 Alice 的 ML-KEM-512 公鑰和 alg 欄位設定為 ML-KEM-512 的物件識別碼。並且本研究主要聚焦在後量子密碼學描述，所以未深入描述序號、效期等資訊，但憑證中應包含這些資訊。另外，由於此憑證主要作為金鑰封裝用，所以在擴展的金鑰用途欄位中加入 dataEncipherment。上述內容作為待簽章憑證，再由憑證中心運用憑證中心 ML-DSA-44 私鑰對為待簽章憑證產製 ML-DSA-44 簽章放到 Sig 欄位。

| Pure Certificate (X.509) |
|---|
| Subject: cn=Alice |
| SPKI: {<br>  alg: **ML-KEM-512 OID**<br>  key: **ML-KEM-512key**<br>} |
| Extensions: |
| Key Usage: {dataEncipherment} |
| Sig: {**ML-DSA-44 signature**} |

**圖 2 支援金鑰封裝用的後量子密碼學 X.509憑證**

## 3. 複合(Composite)混合憑證方案

雖然美國國家標準暨技術研究院於 2024 年 8 月已制定了後量子密碼學標準，但仍有部分專家為提升安全性，提出複合混合憑證方案[13],[14]，讓同一張憑證同時包含有後量子密碼學公鑰和傳統密碼學(包含 RSA 密碼學和橢圓曲線密碼學)公鑰，以及包含後量子密碼學簽章和傳統密碼學簽章。雖然這些技術仍是草案階段，但目前已進入第 6 版以上，未來將可能成為標準。當驗證複合混合憑證方案的憑證時，需要後量子密碼學簽章和傳統密碼學簽章都驗證通過才會認定是合法的憑證。本節案例，假設憑證擁有者是 Alice，並且 Alice 和憑證中心的數位簽章演算法包含有後量子密碼學 ML-DSA-44、橢圓曲線密碼學 ECDSA (Elliptic Curve Digital Signature Algorithm) NIST P-256。

以支援數位簽章用的複合混合憑證方案(如圖 3 所示)為例，在主題的常用名稱欄位中描述 Alice，並且設定主題公鑰資訊的 key 欄位設定為 Alice 的後量子密碼學 ML-DSA-44 公鑰||橢圓曲線密碼學 ECDSA NIST P-256 公鑰、alg 欄位設定為 id-MLDSA44-ECDSA-P256-SHA256 的物件識別碼。另外，由於此憑證主要作為數位簽章用，所以在擴展的金鑰用途欄位中加入 digitalSignature。上述內容作為待簽章憑證，再由憑證中心運用憑證中心後量子密碼學 ML-DSA-44 私鑰和橢圓曲線密碼學 ECDSA NIST P-256 私鑰對為待簽章憑證產製後

量子密碼學 ML-DSA-44 簽章||橢圓曲線密碼學 ECDSA NIST P-256 簽章放到 Sig 欄位。

| Hybrid Certificate (X.509) |
|---|
| Subject: cn=Alice |
| SPKI: {<br>  alg: {<br>      id-MLDSA44-ECDSA-P256-SHA256 OID<br>  }<br>  key: {<br>      ML-DSA-44 key ||<br>      ECDSA P-256 key<br>  }<br>} |
| Extensions: |
| Key Usage: {<br>      digitalSignature<br>} |
| Sig: {<br>    ML-DSA-44 signature ||<br>    ECDSA P-256 signature<br>} |

**圖 3 支援數位簽章用的複合混合憑證方案**

## 4. 催化劑(Catalyst)混合憑證方案

在混合憑證方案設計上,除了第 3 節所的安全性考量外,另外也考量遷移到後量子密碼學過渡時期與傳統密碼學方法憑證的相容性,所以有專家提出催化劑混合憑證方案[15]。讓主題公鑰資訊欄位存放傳統密碼學(包含 RSA 密碼學和橢圓曲線密碼學)公鑰,而 Sig 欄位採用傳統密碼學簽章。然後,再於擴展中加入可供選擇的主題公鑰資訊(Alternative Subject Public Key Information, Alt-SPKI)欄位,用以存放後量子密碼學的物件識別碼、公鑰、簽章[15]。當欲驗證該憑證的設備仍不支援後量子密碼學時,仍可以驗證主題公鑰資訊欄位和 Sig 欄位;當欲驗證該憑證的設備可支援後量子密碼學時,則除了驗證主題公鑰資訊欄位和 Sig 欄位之外,也可以驗證可供選擇的主題公鑰資訊欄位,達到後量子密碼學和傳統密碼學的雙重安全。本節案例,假設憑證擁有者是 Alice,並且 Alice 和憑證中心的數位簽章演算法包含有後量子密碼學 ML-DSA-44、橢圓曲線密碼學 ECDSA NIST P-256。

以支援數位簽章用的催化劑混合憑證方案(如圖 4 所示)為例,在主題的常用名稱欄位中描述 Alice,並且設定主題公鑰資訊的 key 欄位設定為 Alice 的橢圓曲線密碼學 ECDSA NIST P-256 公鑰、alg 欄位設定為橢圓曲線密碼學 ECDSA NIST P-256 的物件識別碼。另外,由於此憑證主要作為數位簽章用,所以在擴展的金鑰用途欄位中加入 digitalSignature。上述內容作為待簽章憑證,再由憑證中心運用憑證中心 ECDSA NIST P-256 私鑰對為待簽章憑證產製 ECDSA NIST P-256 簽章放到 Sig 欄位。除此之外,催化劑混合憑證方案的核心設計在於可供選擇的主題公鑰資訊,所以在可供選擇的主題公鑰資訊中的 key 欄位設定為 Alice 的 ML-DSA-44 公鑰、alg 欄位設定為 ML-DSA-44 的物件識別碼、以及由憑證中心運用憑證中心 ML-DSA-44 私鑰對為待簽章憑證產製 ML-DSA-44 簽章放到 sig 欄位。

值得注意的是,雖然催化劑混合憑證方案是一個可以兼具後量子密碼學公鑰和傳統密碼學安全的混合憑證,並且結構上的設計可以作為遷移到後量子密碼學過渡時期的候選方案之一。然而,該草案目前的狀態已經在 2024 年過期,所以顯示目前國際並沒有認定化劑混合憑證方案為未來的主流方案。

| Hybrid Certificate (X.509) |
|---|
| Subject: cn=Alice |
| SPKI: {<br>    alg: **ECDSA P-256 OID**<br>    key: **ECDSA P-256 key**<br>} |
| Extensions: |
| Key Usage: {<br>      digitalSignature<br>} |
| Alt-SPKI: {<br>    **alg: ML-DSA-44 OID**<br>    **key: ML-DSA-44 key**<br>    **sig: ML-DSA-44 sig.**<br>} |
| Sig: {**ECDSA P-256 signature**} |

**圖 4 支援數位簽章用的催化劑混合憑證方案**

## 5. 變色龍(Chameleon)混合憑證方案

在兼具安全性和可遷移性的混合憑證方案設計上,除了第 4 節所的催化劑混合憑證方案外,另外也有專家提出變色龍混合憑證方案[16],並且變色龍混合憑證方案目前已經進入第 06 版,未來有可能成為標準之一。

變色龍混合憑證方案的核心精神在憑證(以下稱為外部憑證(Outer Certificate))中再加入另一張差異憑證(Delta Certificate)(以下稱為內部憑證(Inner Certificate))在外部憑證的擴展欄位中。其中,在外部憑證主題公鑰資訊欄位存放傳統密碼學(包含 RSA 密碼學和橢圓曲線密碼學)公鑰,而 Sig 欄位採用傳統密碼學簽章。在內部憑證主題公鑰資訊欄位存放後量子密碼學公鑰,而 Sig 欄位採用後量子密碼學簽章。如果外部憑證和內部憑證有相同資訊的欄位(例如:常用名稱欄位、效期等),則在內部憑證欄位可以省略,直接視同該欄位資訊與外部憑證對應的欄位資訊一致[16]。當欲驗證該憑證的設備仍不支援後量子密碼學時,仍可以驗證主題公鑰資訊欄位和 Sig 欄位;當欲驗證該憑證的

設備可支援後量子密碼學時，則除了驗證主題公鑰資訊欄位和 Sig 欄位之外，也可以驗證內部憑證的主題公鑰資訊欄位和 Sig 欄位，達到後量子密碼學和傳統密碼學的雙重安全。本節案例，假設憑證擁有者是 Alice，並且 Alice 和憑證中心的數位簽章演算法包含有後量子密碼學 ML-DSA-44、橢圓曲線密碼學 ECDSA NIST P-256。

以支援數位簽章用的變色龍混合憑證方案(如圖 5 所示)為例，在主題的常用名稱欄位中描述 Alice，並且設定主題公鑰資訊的 key 欄位設定為 Alice 的橢圓曲線密碼學 ECDSA NIST P-256 公鑰、alg 欄位設定為橢圓曲線密碼學 ECDSA NIST P-256 的物件識別碼。另外，由於此憑證主要作為數位簽章用，所以在擴展的金鑰用途欄位中加入 digitalSignature。上述內容作為待簽章憑證，再由憑證中心運用憑證中心 ECDSA NIST P-256 私鑰對為待簽章憑證產製 ECDSA NIST P-256 簽章放到 Sig 欄位。除此之外，變色龍混合憑證方案的核心設計在於內部憑證，所以在內部憑證主題公鑰資訊中的 key 欄位設定為 Alice 的 ML-DSA-44 公鑰、alg 欄位設定為 ML-DSA-44 的物件識別碼，以及由憑證中心運用憑證中心 ML-DSA-44 私鑰對為內部憑證的待簽章憑證產製 ML-DSA-44 簽章放到內部憑證的 Sig 欄位。

```
Hybrid Certificate (X.509)
Subject: cn=Alice
SPKI: {
    alg: ECDSA P-256 OID
    key: ECDSA P-256 key
}
Extensions:
  Key Usage: {
      digitalSignature
  }
  DeltaCertificateDescriptor:
    SPKI: {
        alg: ML-DSA-44 OID
        key: ML-DSA-44 key
    }
    Key Usage: {
        digitalSignature
    }
    Sig: {ML-DSA-44 signature}
Sig: {ECDSA P-256 signature}
```

圖 5 支援數位簽章用的變色龍混合憑證方案

## 6. 混合憑證方案比較與討論

本節將分別從憑證長度、計算時間長度、以及遷移期間過渡方案三個面向來比較複合混合憑證方案、催化劑混合憑證方案、變色龍混合憑證方案，依序於第 6.1 節~第 6.3 節討論。最後，在第 6.4 節總結比較結果。

### 6.1 憑證長度比較

複合混合憑證方案有最短的憑證長度，催化劑混合憑證方案則有次短的憑證長度，而變色龍混合憑證方案的憑證長度最長，比較結果如表 3 所示。

複合混合憑證方案把原本兩個密碼學方法的物件識別碼改為只用一個物件識別碼表述，所以節省了一個物件識別碼。除此之外，複合混合憑證方案把把原本兩個密碼學方法的公鑰合併在一個位元字串(bit string)，所以可以再進一步節省資料長度和型態描述的資訊；在簽章的作法也是相同，所以可以節省資料長度和型態描述的資訊。

表 3 憑證長度比較

| 混合憑證方案 | 憑證長度 |
|---|---|
| 複合混合憑證方案 | 短 |
| 催化劑混合憑證方案 | 短 |
| 變色龍混合憑證方案 | 長 |

催化劑混合憑證方案主要把後量子密碼學的相關資訊放在可供選擇的主題公鑰資訊，而在資料長度和型態描述上則和原本的描述方式一樣，所以比複合混合憑證方案略長一些，但整體方案的憑證長度仍是較短的。

變色龍混合憑證方案擁有最長的憑證長度，原因在於變色龍混合憑證方案主要增加內部憑證，所以等價於兩張憑證的長度。主要在於如果內部憑證和外部憑證有一致資訊時，內部憑證可以省略該欄位，可以比起兩張憑證的憑證長度總和來得少一些，但減少幅度相當有限。

### 6.2 計算時間長度比較

複合混合憑證方案有最短的計算時間，而催化劑混合憑證方案和變色龍混合憑證方案都需要較長的計算時間，比較結果如表 4 所示。

複合混合憑證方案在憑證中心產製簽章時，憑證中心的後量子密碼學 ML-DSA-44 私鑰和憑證中心的橢圓曲線密碼學 ECDSA NIST P-256 私鑰是對相同的待簽章憑證產製簽章，所以實作上可以運用平行計算來加速，讓兩個簽章同時產製。因此，複合混合憑證方案有最短的計算時間。

催化劑混合憑證方案在憑證中心產製簽章時，必須先由憑證中心的後量子密碼學 ML-DSA-44 私鑰對不包含可供選擇的主題公鑰資訊的待簽章憑證產製後量子密碼學 ML-DSA-44 簽章。當產製好後量子密碼學 ML-DSA-44 簽章後，再得到包含可供選擇的主題公鑰資訊的待簽章憑證，然後再用憑證中心的橢圓曲線密碼學 ECDSA NIST P-256 私

鑰是對包含可供選擇的主題公鑰資訊的待簽章憑證產製橢圓曲線密碼學 ECDSA NIST P-256 簽章。因此，後量子密碼學 ML-DSA-44 簽章和橢圓曲線密碼學 ECDSA NIST P-256 簽章必須循序產製，而無法平行計算，所以催化劑混合憑證方案需要較長的計算時間。

表 4 計算時間長度比較

| 混合憑證方案 | 計算時間長度 |
|---|---|
| 複合混合憑證方案 | 短 |
| 催化劑混合憑證方案 | 長 |
| 變色龍混合憑證方案 | 長 |

變色龍混合憑證方案在憑證中心產製簽章時，必須先由憑證中心的後量子密碼學 ML-DSA-44 私鑰對內部憑證的待簽章憑證產製後量子密碼學 ML-DSA-44 簽章。當產製好後量子密碼學 ML-DSA-44 簽章後，再得到包含內部憑證的外部憑證之待簽章憑證，然後再用憑證中心的橢圓曲線密碼學 ECDSA NIST P-256 私鑰是對包含內部憑證的外部憑證之待簽章憑證產製橢圓曲線密碼學 ECDSA NIST P-256 簽章。因此，在變色龍混合憑證方案，後量子密碼學 ML-DSA-44 簽章和橢圓曲線密碼學 ECDSA NIST P-256 簽章也必須循序產製，所以變色龍混合憑證方案需要較長的計算時間。

### 6.3 遷移期間過渡方案比較

催化劑混合憑證方案和變色龍混合憑證方案都能作為後量子密碼學遷移期間過渡方案，但複合混合憑證方案則無法成為後量子密碼學遷移期間過渡方案，比較結果如表 5 所示。

表 5 遷移期間過渡方案比較

| 混合憑證方案 | 是否能成為過渡方案 |
|---|---|
| 複合混合憑證方案 | 否 |
| 催化劑混合憑證方案 | 是 |
| 變色龍混合憑證方案 | 是 |

複合混合憑證方案的主題公鑰資訊欄位包含有後量子密碼學公鑰和傳統密碼學(包含 RSA 密碼學和橢圓曲線密碼學)公鑰，並且 Sig 欄位包含有後量子密碼學簽章和傳統密碼學簽章。當欲驗證該憑證的設備仍不支援後量子密碼學時，則無法驗證該憑證。因此，複合混合憑證方案無法作為後量子密碼學遷移期間過渡方案。

催化劑混合憑證方案的主題公鑰資訊欄位和 Sig 欄位分別僅包含傳統密碼學(包含 RSA 密碼學和橢圓曲線密碼學)公鑰和傳統密碼學簽章。而後量子密碼學的公鑰和簽章則是存放於擴展的可供選擇的主題公鑰資訊。當欲驗證該憑證的設備仍不支援後量子密碼學時，則仍可以驗證主題公鑰資訊欄位和 Sig 欄位，然後可以支援傳統密碼學的安全性；當欲驗證該憑證的設備可支援後量子密碼學時，再驗證擴展的可供選擇的主題公鑰資訊，達到後量子密碼學和傳統密碼學的雙重安全。因此，催化劑混合憑證方案可以作為後量子密碼學遷移期間過渡方案。

變色龍混合憑證方案外部憑證的主題公鑰資訊欄位和 Sig 欄位分別僅包含傳統密碼學(包含 RSA 密碼學和橢圓曲線密碼學)公鑰和傳統密碼學簽章。而後量子密碼學的公鑰和簽章則是存放於內部憑證的主題公鑰資訊欄位和 Sig 欄位。當欲驗證該憑證的設備仍不支援後量子密碼學時，則仍可以驗證外部憑證的主題公鑰資訊欄位和 Sig 欄位，然後可以支援傳統密碼學的安全性；當欲驗證該憑證的設備可支援後量子密碼學時，再驗證內部憑證的主題公鑰資訊欄位和 Sig 欄位，達到後量子密碼學和傳統密碼學的雙重安全。因此，變色龍混合憑證方案可以作為後量子密碼學遷移期間過渡方案。

### 6.4 小結

綜合上述，複合混合憑證方案具有憑證長度短、計算時間短的優勢，所以如果目標是希望建構後量子密碼學和傳統密碼學的雙重安全，則複合混合憑證方案是最佳選擇方案。

催化劑混合憑證方案和變色龍混合憑證方案其優劣較相似，但由於變色龍混合憑證方案的設計上是採用內部憑證的方式，所以可以讓外部憑證和內部憑證的一些欄位(如：金鑰用途)各別表述，所以更彈性。除此之外，催化劑混合憑證方案草案已於 2024 年過期，所以未來可能不會被採用。因此，建議可以選擇變色龍混合憑證方案作為後量子密碼學遷移期間過渡方案。

## 7. 結論與未來發展

本研究主要整理了後量子密碼學、後量子密碼學憑證、以及後量子密碼學混合憑證方案的近期發展。其中，對複合混合憑證方案、催化劑混合憑證方案、變色龍混合憑證方案進行深入的描述和討論，並且從憑證長度、計算時間長度、以及遷移議題三個面向來比較各種混合憑證方案，總結每個混合憑證方案適用的情境，未來可以作為遷移到後量子密碼學的部署參考。

本研究討論的混合憑證方案主要是同樣是數位簽章用的密碼學方法混合或同樣是金鑰封裝用的密碼學方法混合。然而，多用途憑證也是未來的研究方向之一。其中，中華電信在 2025 年參加 NIST Workshop on Guidance for KEMs，並且發表"A Dual-usage Certificate Signing Request Protocol for Digital Signature Algorithm and Key Encapsulation Mechanism"[18]和"Post-Quantum Cryptography-

Based Bidirectional Authentication Key Exchange Protocol and Industry Applications: A Case Study of Instant Messaging"[19]等研究成果，分享多用途憑證的設計原理及其應用。未來可以朝向數位簽章用的密碼學方法和金鑰封裝用的密碼學方法的多用途結合發展。

### 致謝

本研究承蒙美國國家標準暨技術研究院邀請，已於 NIST Workshop on Guidance for KEMs 報告"A Dual-usage Certificate Signing Request Protocol for Digital Signature Algorithm and Key Encapsulation Mechanism"和"Post-Quantum Cryptography-Based Bidirectional Authentication Key Exchange Protocol and Industry Applications: A Case Study of Instant Messaging"等後量子密碼學多用途憑證相關研究成果，特此感謝。